\documentclass[preprint2]{aastex6}

\begin{document}


\title{AKARI near-infrared spectroscopy of the extended green object G318.05+0.09: 
Detection of CO fundamental ro-vibrational emission}


\author{Takashi Onaka, Tamami Mori,
Itsuki Sakon and Aleksandra Ardaseva\altaffilmark{1}}
\affil{Department of Astronomy, Graduate School of Science, The University of Tokyo,
113-0033 Tokyo, Japan}
\email{onaka@astron.s.u-tokyo.ac.jp}

\altaffiltext{1}{School of Physics and Astronomy, University of St. Andrews, 
North Haugh, St Andrews, Fife, KY16 9SS, UK}

\begin{abstract}

We present the results of near-infrared (2.5--5.4\,$\mu$m) long-slit spectroscopy of 
the extended green object (EGO) G318.05+0.09 with {\it AKARI}.
Two distinct sources are found in the slit.  The brighter source has strong red continuum
emission with H$_2$O ice, CO$_2$ ice, and CO gas and ice absorption features at 3.0, 4.25\,$\mu$m,
4.67\,$\mu$m, 
respectively, while the other greenish object
shows peculiar emission that has double peaks at around 4.5 and 4.7\,$\mu$m.
The former source is located close to the ultra compact \ion{H}{2} region
IRAS\ 14498$-$5856 and is identified
as an embedded massive young stellar object (YSO).  The spectrum of the
latter source can be interpreted by blue-shifted ($-3000 \sim -6000$\,km\,s$^{-1}$)
optically thin
emission of the fundamental ro-vibrational transitions ($v=1-0$) of CO molecules with
temperatures of 12000--3700\,K without noticeable H$_2$ and \ion{H}{1} emission.
We discuss the nature of this source in terms of outflow associated with the young
stellar object and supernova ejecta associated with a supernova remnant.

\end{abstract}




\section{Introduction} \label{sec:intro}

The {\it Spitzer} Galactic Legacy Infrared Mid-Plane Survey Extraordinaire
\citep[GLIMPSE:][]{benjamin2003} has discovered a new class of interesting objects
that show extended bright emission at the IRAC band 2 (4.5\,$\mu$m), called
Extended Green Objects \citep[EGOs,][]{cyganowski2008}.  EGOs are often
discussed in terms of candidates of massive young stellar objects (YSOs) with outflows.
The greenish color is attributed either to H$_2$ ($v=$0--0, S(9, 10, 11)) lines and CO ($v=$1--0) 
band heads excited by shocks, or scattered light in the outflow cavity \citep[e.g.,][]{noriega2004, smith2006, cyganowski2008, qiu2008, tobin2008}. 
\citet{takami2010, takami2012} make detailed analysis of the origin of the excess emission based
on photometric data of EGOs and
conclude that extra components other than H$_2$ emission are needed to explain the excess
emission at 4.5\,$\mu$m and that continuum emission of the scattered light in the outflow cavity is likely the primary origin
of the excess, which is supported by \cite{lee2013}.  
Supernova remnants (SNRs) also show excess emission at 4.5\,$\mu$m due to H$_2$ and CO
molecular bands \citep{reach2006}.  While H$_2$ emission is thought to dominate
in the line emission, the contribution from
CO emission becomes comparable to that from H$_2$ for the H$_2$ density of a range 
$10^6-10^7$\,cm$^{-3}$ \citep{neufeld2008}.  CO ($v=1-0$) emission is not detected in
the SNR IC443 by ground-based observations \citep{richter1995}, but it is detected toward the
Orion molecular cloud \citep[OMC][]{geballe1987, geballe1990, gonzalez2002}, suggesting that CO emission
could make a contribution to the excess emission.  

It is indispensable to make spectroscopic observations in order to study the
origin of the excess at 4.5\,$\mu$m in EGOs unambiguously.
However, no direct spectroscopy has so far been made for EGOs except for
the 2\,$\mu$m region \citep{caratti2015} due to the lack of sensitive spectrometer in 
the 4\,$\mu$m region.  In {\it this paper}, we report the result of spectroscopy in 2.0--5.3\,$\mu$m
of one of the
EGOs, G318.05+0.09, made with the Infrared Camera (IRC) onboard {\it AKARI}, which
offered high-sensitivity spectroscopy in 2.0--5.3\,$\mu$m even in its warm mission phase
\citep{onaka2010}.

\section{Observations and data reduction} \label{sec:obs}
\begin{figure}[ht!]
\plotone{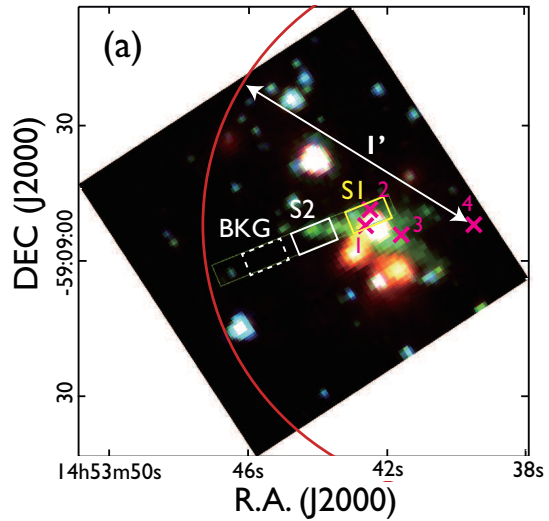}
\caption{{\it Spitzer}/IRAC artificial three-color image of the region of EGO G318.05+0.09 and the slit
position of the present {\it AKARI}/IRC spectroscopy (blue: IRAC band 1 (3.6\,$\mu$m), green:
IRAC band 2 (4.5\,$\mu$m) and red: IRAC band 3 (5.8\,$\mu$m)).  
The thin green rectangle indicates the location of the IRC Ns
slit ($5\arcsec \times 40\arcsec$).  
The western yellow box, the center white box, and the eastern white dotted box in the slit indicate the regions for which the spectra of $S1$, $S2$ and the background
emission are extracted, respectively.  
The yellow box for which the spectrum of $S1$ is extracted includes part of the bright source.
The crosses indicate the positions of the known sources in the catalogs (Table~\ref{tab1}).  The sources
are labeled by the numbers: 1 is the methanol maser source
\citep{caswell2009, green2012}, 2 EGO G319.05+0.09 \citep{cyganowski2008}, 3
IRAS\ 14498--5856, and 4 Source B of {\it BeppoSAX} observations \citep{bocchino2001}.  The red circle shows its error circle
of the X-ray source position ($1\arcmin$).
The IRAC data are
retrieved from the NASA/IPAC Infrared Science Archive.
\label{fig1}}
\end{figure}

The {\it AKARI} /IRC observation of G318.05+0.09 was carried out in the director time on 2010 February 18, which was one of the last observations of the {\it AKARI}  mission
(Observation I.D.: 5201570.1). 
The observation was made with the slit spectroscopy mode at the Ns slit of $5\arcsec$ width with the 
prism and grism dispersers (AOT IRCZ4 with c;Ns).  
The prism spectroscopy provides spectra of 2.0--5.3\,$\mu$m with the spectral resolution of
about 0.14\,$\mu$m at 4\,$\mu$m, while spectra of 2.5--5.0\,$\mu$m with
the resolution of about 0.03\,$\mu$m are obtained with the grism.
In the present observation, four exposure frames with 
the prism were first taken and then five exposures with grism were performed, 
which sandwiched a 3.2\,$\mu$m-band imaging observation
in-between 
that was used for the position determination \citep[see][for more details of the IRC
spectrsocpy]{onaka2007, ohyama2007}.
It provided long slit spectroscopy of about $40\arcsec$ long
along the slit direction.  The slit location is shown in Figure~\ref{fig1} 
superimposed on the {\it Spitzer}/IRAC artificial three-color image together with the 
positions of the known sources
in the catalogs (see Table~\ref{tab1}).  The slit of the present observation
runs from an infrared bright region at the
western edge, which is most likely to be IRAS\ 14498--5856, although the catalog position is slightly
different from the bright source (cross \#3 in Fig.~\ref{fig1}).  
The slit covers a greenish blob at around its center.  
The two distinct sources are clearly seen in the long-slit spectrum image.  We call the bright source
$S1$ and  the greenish blob $S2$ in the following.
Spectra are extracted from $5\arcsec \times 8.7\arcsec$ (6 pixels in the slit direction) areas
indicated by the yellow and white boxes both for the grism and prism data.  
The former corresponds to $S1$ and the latter to $S2$.
The slit area of $S1$ does not collect the entire emission from the bright source.  
EGO G318.05+0.09 is defined by a diamond-like shape 
area of $\sim 35\arcsec \times 10\arcsec$, whose center
is located at slightly north of the bright source as marked by cross \#2.  It 
extends to the east and west directions symmetrically to enclose the bright source and
cover the greenish regions, including $S2$ \citep[Fig.~3.14 of][]{cyganowski2008}.  
The present {\it AKARI}/IRC spectroscopy covers
only the eastern part of G318.05+0.09.   
The center position of EGO G318.05+0.09 is located in the region for which the spectrum of
$S1$ is extracted.  The methanol maser source MMB\,318.050+0.087 \citep{caswell2009,
green2012} is also located within the box of $S1$ (cross \#1).
Table~\ref{tab1} summarizes the central positions of the regions where the spectra are
extracted together with the known sources in this region.

The data reduction was performed with the latest toolkit for the phase 3 (warm mission phase)
of the IRC spectroscopy, 
IRC\_SPEC\_TOOLKIT\_P3\_20150331\footnote{\url{http://www.ir.isas.jaxa.jp/AKARI/Observation/support/IRC/}}.  
Since the observation was made
in the very last period of the warm mission phase of {\it AKARI}, the effect of hot pixels 
became severe and the noise level was higher than typical values in the 
phase 3 spectroscopy \citep{onaka2010}.  
To remove residual dark current and possible contribution from the
surrounding background/foreground emission, we subtract the emission enclosed by 
the white rectangle in the eastern part of the slit labeled by BKG in Figure~\ref{fig1}.
There is a shift in the slit position on the detector array between the grism and prism spectroscopy
due to the optical alignment.
The amount of the shift is estimated by matching the bright region and the spectra of the same region are extracted within an accuracy of
a half pixel ($\sim 0.8\arcsec$) along the slit direction for the grism and prism modes.
Details of the data reduction are summarized in \citet{mori2014}.
Due to the relatively large number of hot pixels, a few residual hot pixels remain after the standard
reduction process.  They are masked manually.  The masked pixels are at 4.457\,$\mu$m and 4.5154\,$\mu$m of the grism spectrum
of $S1$ and at 3.9607--3.9801\,$\mu$m in the grism spectrum of $S2$ and they do not affect the following discussion.
It should also be noted that the absolute calibration may have a large uncertainty ($\sim 20$\%) 
due to the
elevated temperature of the detector array at the observing time ($\sim 47.5$\,K)
because the sensitivity is not well calibrated at warm temperatures, while no change in the spectral response due to the change in the temperature is recognized
\citep{onaka2010}.
We have not applied correction for the contamination of the second order light for the grism
spectra in the phase 2 (cold mission phase) 
recently reported by \citet{baba2016} because there is no calibration of the contamination estimate
available for the phase 3 data at present.  
Instead we discard the data in the spectral region where the contamination is expected
($> 4.9$\,$\mu$m).  The improved wavelength calibration provided by \citet{baba2016} is
applied to the present grism spectra. 
The correction for wavelengths longer than 4\,$\mu$m is not significant ($<0.006$\,$\mu$m) compared
to the instrumental spectral resolution.

\begin{deluxetable*}{clllc}
\tablecaption{Positions of the sources in the G318.05+0.09 region\label{tab1}}
\tablecolumns{5}
\tablehead{
\colhead{Source name} &
\colhead{R.A. (J2000)} &
\colhead{Decl. (J2000)} &
\colhead{Note} &
\colhead{Number in Fig.~\ref{fig1}}
}
\startdata
$S1$ & 14 53 42.2 & -59 08 48.8 & Present work\tablenotemark{a} \\
$S2$ & 14 53 44.0 & -59 08 54.3 & Present work\tablenotemark{a}  \\
MMB318.050+0.087 & 14 53 42.67 & -50 08 52.4 & Methanol maser\tablenotemark{b} & 1\\
G318.05+0.09 & 14 53 42.6  & -59 08 49\quad & EGO ($\sim 35\arcsec \times 10\arcsec$)\tablenotemark{c} & 2\\
IRAS\ 14498--5856 & 14 53 41.6  & -59 08 54 & UC\ion{H}{2}\tablenotemark{d}  & 3\\
Source B & 14 53 39.5 & -59 08 52 & X-ray source\tablenotemark{e} & 4\\
G318.2+0.1 & 14 54 50 & -59 46    & SNR\tablenotemark{f}  & outside of Fig.~\ref{fig1}
\enddata
\tablenotetext{a}{The center position from which the spectrum is extracted in the slit.}
\tablenotetext{b}{\citet{caswell2009, green2012}}
\tablenotetext{c}{\citet{cyganowski2008}}
\tablenotetext{d}{\citet{walsh1997}}
\tablenotetext{e}{\citet{bocchino2001}}
\tablenotetext{f}{\cite{whiteoak1996}}
\end{deluxetable*}

\section{Results} \label{sec:results}
\begin{figure*}[ht!]
\plotone{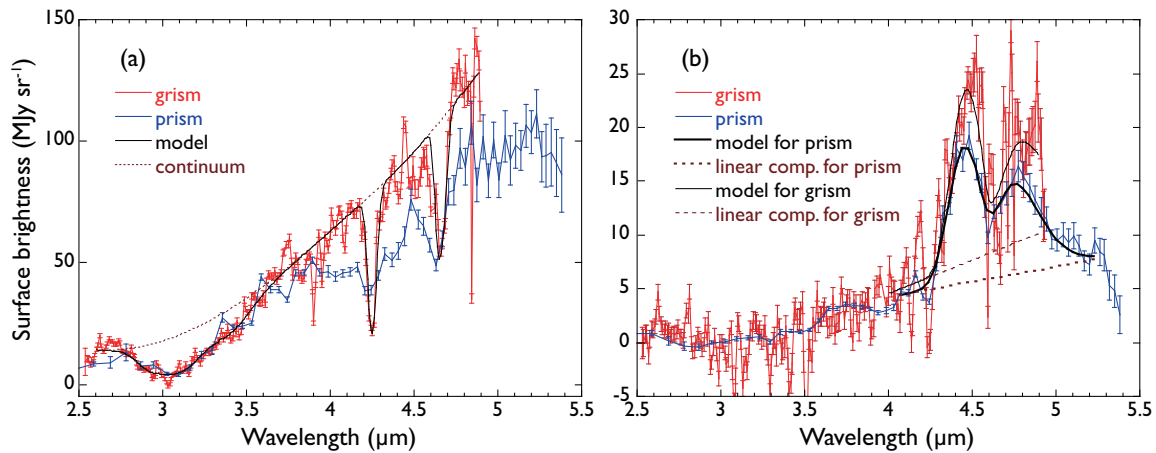}
\caption{{\it AKARI}/IRC near-infrared spectra of (a) $S1$ and (b) $S2$.  
The red lines indicate spectra taken with grism, while the blue lines show those with prism.
In (a) the model spectrum with ice and gas absorption features and the assumed continuum
are shown by the solid black and dotted brown lines, respectively, for the grism spectrum.
In (b) the model spectra with CO emission with linearly increasing components
are shown by the thick and thin black lines for the prism and grsim spectra, respectively.
The assumed linear component is shown by the thick and thin brown dotted lines for the prism
and grism spectra, respectively.  For the prism spectrum, the 
CO emission is assumed to have the temperature
of 1600\,K
and the velocity of $-4700$\,km\,s$^{-1}$, while they are 
assumed to be 2000\,K and $-4100$\,km\,s$^{-1}$
for the grism spectrum (see text for details). 
\label{fig2}}
\end{figure*}

Figure~\ref{fig2} shows the extracted spectra of $S1$ and $S2$.  There is apparent discrepancy
between the grism and prism spectra of $S1$ in 4--5\,$\mu$m.  We attribute the difference to the
saturation effect of the prism spectrum since $S1$ is very bright for the prism spectroscopy
in phase 3.   The grism spectrum does not have a saturation problem at this brightness level
and thus we regard it more reliable.  The grism spectrum shows absorption features
at around 3.0, 4.25, and 4.67\,$\mu$m clearly, whose traces are also seen in the prism spectrum.
The former two can be attributed to H$_2$O and CO$_2$ ice absorption, and the latter to
CO gas or ice absorption.  
To estimate the absorption features quantitatively, 
we fit the continuum by a spline curve with anchor points
at 2.6, 2.65, 3.8, 4.1, 4.4, and 4.9\,$\mu$m (thin brown dotted line in Fig.~\ref{fig2}).  

\begin{figure*}[ht!]
\plotone{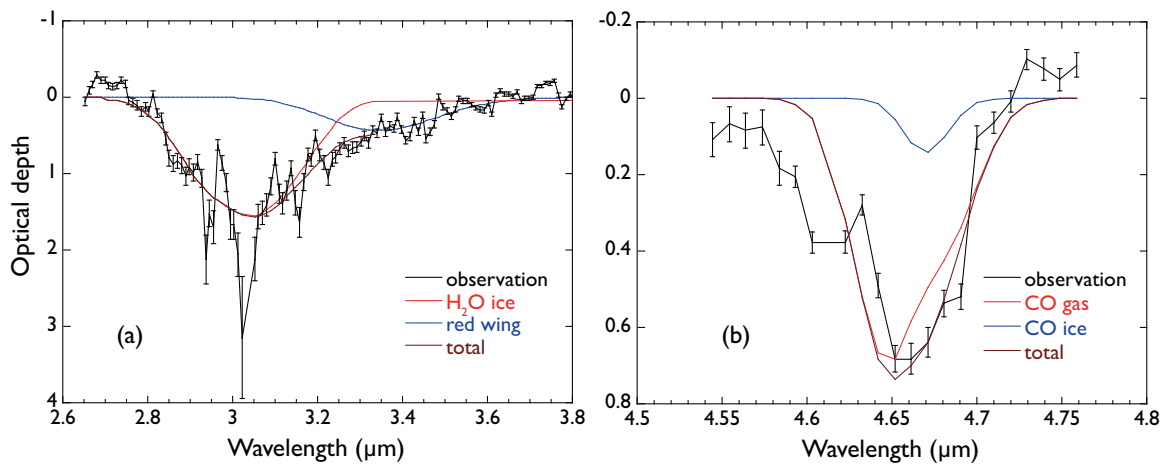}
\caption{Absorption features and model fits in units of the optical depth for the grism spectrum of
$S1$. (a) 3\,$\mu$m region
and (b) 4.67\,$\mu$m region. (a)
The black line shows the observation, the red line the H$_2$O ice component,
the blue line the red wing component approximated by a Gaussian function, and the
brown line the summation of H$_2$O ice and the red wing. 
(b) The black line shows the observation, the red line the CO gas component of 20K,
the blue line the CO ice component approximated by a Gaussian, and the
brown line the summation of the CO gas and ice features. 
\label{fig3}}
\end{figure*}

For
the H$_2$O ice absorption profile, the laboratory data taken at 10\,K are adopted \citep{ehrenfreund96}.
We find that an extra ``red wing'' component is
needed to fit the 3\,$\mu$m absorption feature in addition to the laboratory data profile.  The red wing is
seen in many objects, but its origin is not yet fully understood \citep[e.g.,][]{smith1993, thi2006, noble2013}.  In the present fit, it is approximated by a
Gaussian profile with the central wavelength of
3.35\,$\mu$m and the FWHM of 0.31\,$\mu$m.  
Figure~\ref{fig3}a shows details of the
fit of the 3\,$\mu$m absorption feature in units of the optical depth.

We assume a Gaussian function for the CO$_2$ ice feature at 4.25\,$\mu$m because it
is narrow compared to the present spectral resolution
and use the band strengths of CO$_2$ in \citet{gerakines1995} to estimate the column density.
The observed profile is well fitted by a Gaussian with 
the central wavelength of 4.25\,$\mu$m and the FWHM of 0.036\,$\mu$m. 
The FWHM is adopted from the 
study of {\it AKARI}/IRC Ns spectroscopy of Galactic star-forming regions \citep{mori2014}.
The central wavelength of the absorption peak of CO$_2$ ice
is slightly shorter than the typical value of 4.26\,$\mu$m obtained
in the same study, but it is
still within the uncertainty due to the large noise level and the spectral resolution of the
present observation.  
The absorption feature at around 4.67\,$\mu$m is broader than the CO ice feature
\citep{gibb2004, mori2014}.  We fit it with both CO gas and ice components.  For CO ice,
we adopt the central wavelength of 4.67\,$\mu$m and the FWHM of 0.031\,$\mu$m obtained
from the Ns spectroscopy by \citet{mori2014} and assume the band strength reported 
in \citet{gerakines1995}.
The observed feature can be fitted by absorption of CO gas at 20K and CO ice
as shown in Fig.~\ref{fig3}b.  The contribution from CO gas dominates in the absorption.
With the present spectral resolution, the P and R branches of CO gas absorption
are not resolved.  
The red side is relatively well fitted, while the fit is not good at the blue side,
where there might be an additional component at around 4.62\,$\mu$m.
Because of the large noise, however, it is difficult to make further discussion on the additional 
component.
CO gas with 15--30\,K gives similar fits and 
the gas temperature cannot be well constrained.   
As Fig.~\ref{fig3} suggests, the contribution from CO ice also has
a large uncertainty.  Therefore, the present estimate of the column densities of CO gas and ice
should be taken with caution.
The derived column densities are $(26.3 \pm 0.5)$, $(3.42\pm 0.03)$, $(2.0 \pm 0.5)$, and
($62 \pm 2)\times 10^{17}$\,cm$^{-2}$ for H$_2$O ice CO$_2$ ice, CO ice, and CO gas, respectively.

Source $S2$ shows a peculiar spectrum.  It does not show emission
features at 3.3--3.5\,$\mu$m, which are generally seen in star-forming regions
and are attributed to polycyclic aromatic hydrocarbons \citep{mori2014},
nor ice absorption features seen in YSOs as in $S1$.  It shows two broad peaks
at around 4.5 and 4.7\,$\mu$m and no significant emission is seen below 4\,$\mu$m.
Both the prism and grim spectra show similar characteristics and thus we conclude that
these are real features despite the noisy spectra. 
A small difference between the grism and prism spectra
is seen at wavelengths longer than 4.5\,$\mu$m.  It can be attributed partly to
the large noise in the grism spectrum and/or to a possible difference in the spectrum
extracted region due to the shift of the slit positions on the detector array (see \S\ref{sec:obs}).

The $S2$ spectrum can be accounted for by blue-shifted optically thin emission of the fundamental
ro-vibrational transitions ($v=1-0$) of CO molecules with red continuum.  
The two peaks are attributed to the P and R
branches of CO gas emission.  Since the grism spectrum is very noisy and does not
have data longer than 4.9\,$\mu$m, we use the prism spectrum to estimate the
temperature and the velocity of the CO gas.  We simply assume that the continuum increases
linearly with the wavelength and the 
CO emission is approximated by the local thermodynamic equilibrium (LTE) condition.  From the fit, the temperature and the velocity 
of the CO gas are estimated as $1600^{+500}_{-400}$\,K and 
$-4700^{+1200}_{-1500}$\,km\,s$^{-1}$, respectively.  
The grism spectrum requires more steeply increasing continuum, but the best fit parameters of the
CO gas (2000\,K and $-4100$\,km\,s$^{-1}$) are within the uncertainties estimated from 
the prism spectrum. 
The model spectra with the best fit parameters are shown in Fig.~\ref{fig2}b both for the
prism (thick black line) and grism spectra (thin black line) together with the linear continuum (thick and
thin dotted brown lines).
The column density of the CO gas is estimated from the prism spectrum with the best fit parameters
as $(1.2\pm0.1)\times 10^{10}$\,cm$^{-2}$, which does not include the systematic
error due to the calibration uncertainty at the time of the observation (\S\ref{sec:obs}).

The shift in the wavelength corresponding to the estimated blue-shift velocity is $-0.06 \sim -0.07$\,$\mu$m, which is about a half of
the resolution of the prism spectrum, but more than twice of that of the grism spectrum at 4.5\,$\mu$m.
The latest wavelength calibration of the grism spectroscopy has been applied \citep[\S\ref{sec:obs},][]{baba2016} and
the accuracy is estimated to be better than 0.006\,$\mu$m, which is much smaller than the shift.
The wavelength calibration of the prism spectroscopy is less accurate and estimated to be better than
1 pixel or 0.04\,$\mu$m at 4--5\,$\mu$m \citep{onaka2009}.  
The comparison of the prism and grism spectroscopy confirms that both wavelengths match with each
other within this accuracy \citep{mori2014}.

\begin{figure}[h!]
\plotone{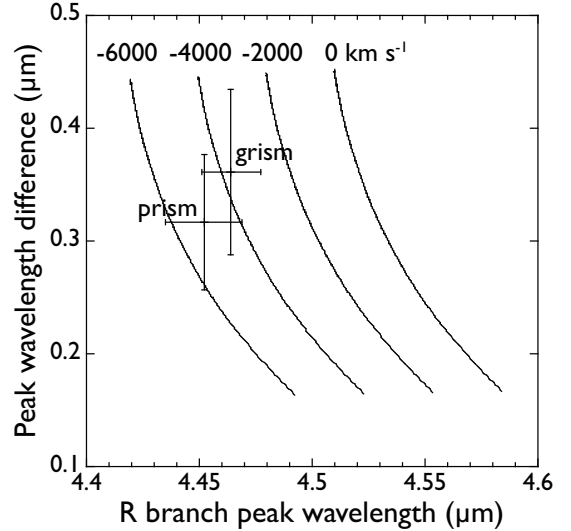}
\caption{Difference in the peak wavelengths of the R and P branches of CO fundamental ro-vibrational emission
are plotted against the peak wavelength of the R branch.
Those from the observed $S2$ spectra with the prism and grism are shown by the crosses with the error bars.
The error bars include the uncertainties in the Gaussian fit and the effect of 
the underlying continuum slope.  The former uncertainties dominate in the errors.
The solid lines indicate the values estimated from the LTE CO emission for temperatures of 500--4000\,K
with the different blue-shift velocities labeled at the top of the lines.  The bottom of the lines corresponds to $T=500$\,K
and the top to $T=4000$\,K.  
\label{fig4}}
\end{figure}

The shift is constrained by the profile shape of the CO emission in the fit and non-LTE effects could affect
the shift estimate.  To investigate the reliability of the suggested large blue-shift velocity, 
we simply calculate the peak positions of the observed spectra
and compare them with those of the LTE model spectra of different temperatures.  The peak positions are estimated 
by a Gaussian fit in the ranges 4.3--4.6\,$\mu$m and 4.65--4.9\,$\mu$m for the R and P branches, respectively.
We also vary the slope of the underlying continuum for 2--7\,MJy\,sr$^{-1}$\,$\mu$m$^{-1}$, which is determined by
the range of the slopes obtained in the CO profile fit.  The effect of the continuum 
slope is found to be smaller ($<0.005$ and $< 0.01$\,$\mu$m for the R and P branches, respectively) 
than the uncertainties in the Gaussian fits both
for the prism and grism spectra.  The peaks of the model CO emission are estimated in the same manner.

Figure~\ref{fig4} plots the difference between the P and R branch
peak wavelengths against the estimated peak wavelength of the R branch 
both for the prism and grism spectra together with those of the model CO emission for the
temperature range 500--4000\,K with the various blue-shift velocities.  
The peak wavelength is sensitive both to the temperature and the blue-shift velocity, but
the difference in the branch peaks is sensitive mostly to the temperature.   The latter is used
to separate the temperature effect from the velocity effect.
The peak wavelength of the R branch of the prism spectrum agrees with that of the grism spectrum within
the calibration uncertainty of the prism spectroscopy.
The prism and grism spectra show compatible values for
the difference in the R and P branch peak wavelengths within the uncertainties in the fit.  Both data points 
suggest that the observed spectra can be accounted for by CO emission with blue-shift velocities of
$-3000 \sim -6000$\,km\,$^{-1}$ and that it is difficult to explain the observed peak wavelengths by simple LTE models
with velocities larger than $-2000$\,km\,s$^{-1}$.
The suggested temperatures of the CO emission range from 1200 to 3700\,K, which are in agreement with the
CO emission profile fit.  While these analyses suggest a large blue-shift velocity consistently, the present
spectra do not have a high signal-to-noise ratio and good spectral resolution.  
Spectra with higher spectral resolution and better quality are needed to investigate the non-LTE effects and
confirm the large blue-shift velocity.

\section{Discussion} \label{sec:discussion}
\subsection{Nature of S1}
Source $S1$ is located close to the position of IRAS\ 14498$-$5856.  The IRAS source is classified
as an ultra compact \ion{H}{2} (UC\ion{H}{2}) region based on the $IRAS$ colors and  methanol maser emission at 6668 MHz is detected toward this object \citep{walsh1997}.  The position of the maser
is accurately measured by \citet{caswell2009}, which matches well with the position
of $S1$ (Fig.~\ref{fig1} and Table~\ref{tab1}).  The methanol maser at 6668 MHz is known as a good tracer
for the early stages of high-mass star formation \citep[and references therein]{green2012}.
The center position of the EGO\,G318.05+0.059 is close to the maser position and it is suggested
to be a likely massive YSO outflow candidate in the EGO catalog \citep{cyganowski2008}.
The {\it AKARI}/IRC spectrum of $S1$ shows the absorption features of H$_2$O and
CO$_2$ ices clearly.  These ice absorption features are also characteristics of 
embedded YSOs \citep[e.g.,][]{gibb2004, shimonishi2010}.  
The column densities of CO$_2$ and H$_2$O ices are known to show a correlation in
YSOs and the ratio depends on the nature of the object 
\citep{gerakines1999, nummelin2001, pontoppidan2008}.
The abundance ratio of CO$_2$ to H$_2$O ices in  
$S1$ is $13.0\pm0.3$\%, being within the range of
the ratios found for weakly processed ices in massive YSOs in our Galaxy \citep[13--23\%,][]
{gibb2004}.  
The CO ice absorption is generally not abundant.  While the relative abundance of CO ice to CO$_2$ ice
varies with the line of sight \citep{pontoppidan2008, noble2013},
the present relative abundance ($7.6\pm1.9$\% relative to H$_2$O ice)
is also within the range found for weakly processed ices in massive YSOs \citep[3--8\%,][]{gibb2004}.
All of these pieces of evidence suggest that $S1$ is most likely an embedded massive YSO.
\citet{walsh1997} estimate a kinematic distance to the maser source as 4.3 or 11.1\,kpc
from its velocity ($-46$\,km\,s$^{-1}$).  An X-ray source has also been detected  by 
{\it BeppoSAX} in this region
\citep[Source B in][]{bocchino2001}.  Source $S1$ is  located within the error circle of the
position of the X-ray source (\#4 in Fig.~\ref{fig1}).  \citet{bocchino2001} suggest that
if the X-ray source is the hard X-ray counterpart of the UC\ion{H}{2},
the shorter distance is favored since the X-ray luminosity becomes too high as an UC\ion{H}{2}
at 11.1\,kpc .
%

The area from which the spectrum of $S1$ is extracted includes greenish emission
as Fig.~\ref{fig1} suggests and thus the spectrum may have a contribution from CO emission.
The prism spectrum of $S1$ suggests
a peak at around 4.5\,$\mu$m.  It may mimic the R branch of CO emission seen in the spectrum of
$S2$.
However, the prism spectrum does not show a P-branch peak at around 4.7\,$\mu$m 
clearly as seen in $S2$.  
Therefore, 
we conclude that the peak at around 4.5\,$\mu$m is a result of a combination of the saturation
effect and the cold CO gas absorption and that there is no clear evidence for warm CO emission
in the spectrum of $S1$.

\subsection{Nature of S2}
Source $S2$ shows a quite distinct spectrum.  It can be interpreted in terms of
highly blue-shifted CO emission with red continuum.  
In the following, we discuss two possibilities for its nature: (1) outflow associated with
the YSO and (2) supernova ejecta associated with a SNR.

Excess emission at 4.5\,$\mu$m in EGOs has often been discussed in terms of outflows 
associated with massive YSOs. Since $S1$ is most likely a massive YSO, $S2$ could be related to
outflow associated with $S1$.  Observations with the Short-Wavelength Spectrometer (SWS)
on board the Infrared Space Observatory ({\it ISO}) detect the fundamental ro-vibrational
($v=1-0$) emission of CO molecules toward the OMC-1 outflow region
together with emission of \ion{H}{1} recombination
lines and H$_2$ lines \citep{rosenthal2000}.  
The SWS observations of OMC-1 resolve each transition of the CO emission and
thus the physical conditions of the CO gas and excitation mechanism are investigated in
detail \citep{gonzalez2002}.  The observed spectra show stronger emission of the P branch than that of the R branch.
\citet{gonzalez2002} conclude that the CO emission consists of warm ($T = 200-400$\,K)
and hot ($T \sim 3000$\,K) components and the P-R-asymmetry can be accounted for by a large CO 
column density ($\sim 6\times 10^{18}$\,cm$^{-2}$).  Detection of the CO ro-vibrational emission toward OMC-1 supports
the possible association of $S2$ with outflow from $S1$.  However, the characteristics of
OMC-1 CO emission are different from the spectra of $S2$. 
There seem to be no 
\ion{H}{1} recombination lines and H$_2$ lines in the grism spectrum of $S2$, which are seen in OMC-1.   
There is excess emission at around the position of 0-0 S(11)  (4.1810\,$\mu$m) in the grism spectrum of $S2$,
but the absence of similar excess emission at 0-0 S(9) (4.6847\,$\mu$m), 0-0 S(10) (4.4096\,$\mu$m), 
and 0-0 S(13)  (3.8464\,$\mu$m)
suggests that the excess at $\sim 4.2$\,$\mu$m is not H$_2$ 0-0 S(11) emission.

To discuss the implication of non-detection of H$_2$ line emission in $S2$,
we make a simple comparison with the spectrum of OMC-1.
The total CO $v=1-0$ emission flux at Peak 1 of OMC-1 is estimated as $3.8 \times 10^{-13}$\,W\,m$^{-2}$
and that of the hot component is  $1.7 \times 10^{-13}$\,W\,m$^{-2}$ \citep{gonzalez2002}.
The fluxes of the four strongest lines of H$_2$ at Peak 1 in 3.5--5\,$\mu$m, 
0-0 S(9), 0-0 S(10), 0-0 S(11), and 0-0 S(13),
are estimated as 3.2, 0.83, 1.4, and $0.92 \times 10^{-14}$\,W\,m$^{-2}$,
respectively \citep{rosenthal2000}.  
The total CO $v=1-0$ emission intensity from $S2$ is estimated as  $1.75 \times 10^{-6}$\,W\,m$^{-2}$\,sr$^{-1}$.
If the same ratios of the H$_2$ line intensities to the total CO emission as in OMC-1 are assumed,
the expected H$_2$ line intensities in $S2$ are 15, 3.8, 6.5, and $4.2 \times 10^{-8}$
\,W\,m$^{-2}$\,sr$^{-1}$ for 0-0 S(9), 0-0 S(10), 0-0 S(11), and 0-0 S(13), respectively.  
If the H$_2$ emission is associated only with the hot CO component in OMC-1,
the expected intensities are 34, 8.5, 14, and $9.4 \times 10^{-8}$
\,W\,m$^{-2}$\,sr$^{-1}$ for 0-0 S(9), 0-0 S(10), 0-0 S(11), and 0-0 S(13), respectively.  
Three-$\sigma$ upper limits of the line intensity in
the $S2$ grism spectrum are estimated as 1.8, 1.7, 4.5, and $1.6 \times 10^{-8}$\,W\,m$^{-2}$\,sr$^{-1}$, respectively,
for 0-0 S(9), 0-0 S(10), 0-0 S(11), and 0-0 S(13).  The upper limit on 0-0 S(11) is relatively large compared to the others
due to the excess emission described above.  All of these upper limits are smaller than the expected intensities
estimated from the spectrum of Peak 1 of OMC-1.
Thus, the H$_2$ line emission relative to the CO emission in $S2$ are weak 
compared to that in OMC-1.  
\citet{gonzalez2002} show that a high H$_2$ density ($10^7$\,cm$^{-3}$) is required to account for
 the observed flux of the hot component of CO emission in OMC-1.  They also find
 that the derived CO column density of the hot component ($\sim 1.8 \times 10^{17}$\,cm$^{-2}$) is much larger than 
 that expected from the column density of hot H$_2$ component \citep[$2.4 \times 10^{19}$\,cm$^{-2}$,][]{rosenthal2000},
 suggesting partial dissociation of H$_2$ in the region.
 \citet{neufeld2008} also show that CO emission becomes comparable to H$_2$
emission in IRAC band 2 only at high gas densities ($10^6 - 10^7$\,cm$^{-3}$).
Non-detection of H$_2$ line emission in $S2$ may thus suggest even more H$_2$-poor conditions
than in OMC-1.  
A detailed study of the physical conditions of CO and H$_2$ in $S2$ is definitely needed by observations
with higher spectral resolution to make quantitative discussion.
It should be noted that the suggested large blue-shift velocity, which also has to be confirmed by
further observations, seems to be 
difficult to be explained if the CO emission of $S2$ originates from outflow associated with $S1$.

Spectra of $S2$ show strong similarity to those detected toward ejecta knots in the SNR Cas A
\citep{rho2012}.  The suggested blue-shift velocity of $S2$ is in the range of those
found in ejecta of Cas A ($-340 \sim -5660$\,km\,s$^{-1}$) and is
attributed to the ejecta motion.  The Cas A spectra 
do not show \ion{H}{1} recombination lines and H$_2$ emission lines either, which can be
attributed to hydrogen-poor abundance of the ejecta.

EGO G318.05+0.09 is located within the SNR G318.2+0.1, which is recognized by a
large ($40\arcmin \times 35\arcmin$) radio shell \citep{whiteoak1996}.  
However, there is no direct evidence that $S2$ is associated with SN ejecta.  
No associated infrared emission has been found for this SNR region by the study of
the GLIMPSE survey \citep{reach2006}.  The present EGO is
a very tiny portion of the entire SNR G318.2+0.1 and this is very different from Cas A, in which extended
infrared emission associated with SNR ejecta features is clearly detected. 
Toward the region of the SNR G318.2+0.1, only {\it BepposSAX} observations are available in the X-ray
\citep{bocchino2001},
which have a large error circle (Fig,~\ref{fig1}) and do not provide clear evidence for the association of
$S2$ with SN ejecta. 
While the characteristics of the spectra of $S2$ suggest emission from
SN ejecta, there is no other supporting evidence for the identification as ejecta
at present.

If $S2$ found in the present observation is truly SN ejecta, this will be the second SNR that
shows CO emission at 4\,$\mu$m.  
SNe are thought to be the major dust suppliers in the interstellar space as well as in the early Universe
\citep{gall2011}.  
However, mid-infrared observations of SNe 
so far suggest that the mass of the warm dust associated
with SNe falls below $10^{-2}$\,$M_\odot$, significantly lower than theoretical predictions 
\cite[references therein]{tinyanont2016}.
Recent far-infrared observations detect a significant amount ($>$0.1\,$M_\odot$) of cold dust
associated with a SN or a SNR \citep{barlow2010, sibthorpe2010, matsuura2011, gomez2012, temim2013, indebetouw2014, 
lau2015, matsuura2015}.  
Detection of SNe and SNRs in the far-infrared is, however, 
limited to a small number of objects and the dust production yield in SNe still remains under debate
\citep[e.g.,][]{gall2011}.    
Recent observations also reveal the presence of several molecules in SNe or SNRs, such as
SiO \citep{kotak2006}, CO \citep{kamenetzky2013}, and ArH$^+$ \citep{barlow2013}.
The presence of CO gas is of particular interest  
since CO traps the major dust-constituting element carbon and may reduce the
carbon dust formation yield significantly \citep{clayton2001}.
CO molecules have been detected in about 10 SNe within 340 years after explosion.  
Cas A is a SNR of 340 years old and it is the only SNR in which both high-J rotational 
and ro-vibrational lines of warm CO gas
have been detected \citep{rho2012, wallstrom2013}. 
Processing and evolution of dust and gas in SN ejecta after the explosion,
in particular the effects of the reverse shock, have been
becoming an important issue for the understanding of the net dust formation yield in SNe
\citep{bianchi2007, nozawa2007, nozawa2010,nath2008, silvia2010, silvia2012,
gall2014, dwek2015, wesson2015, biscaro2016}.
\citet{biscaro2014} show that
dust and molecules in ejecta clumps are destroyed by the reverse shock, whereas CO gas can reform 
in the post-reverse shock gas, but dust clusters do not.  
Processing of dust and gas in the post-reverse shock clumps depends sensitively on
the environmental conditions \citep{biscaro2016}.
While no information is available of the SNR with which $S2$ is associated, if it is confirmed to be SN ejecta,
further studies of $S$ will 
provide important information on the study of the physical and chemical conditions of the ejecta.

The presently available data for EGO G318.05+0.09 are not sufficient to draw a clear conclusion
on the nature of $S2$.   Spectroscopy with high spectral resolution and good quality are
definitely needed to study the physical conditions of the CO gas and to confirm the large blue-shift
velocity.  Future observations with {\it JWST}/NIRSpec will certainly give us crucial information for the identification.
X-ray observations with high spatial resolution and good energy resolution of 
{\it Chandra} will also help identify the nature of $S2$.

\section{Summary}
We carried out near-infrared spectroscopy of the EGO G318.05+0.09 with {\it AKARI}/IRC.
We have detected two distinct infrared sources in the region.  
One, $S1$, is associated with the UC\ion{H}{2}
IRAS\ 14498$-$5856 and identified as an embedded massive YSO based on the presence of the
absorption features due to H$_2$O and CO$_2$ ice species together with the associated
methanol maser emission.  The other greenish-color
source, $S2$, shows a quite peculiar spectrum, which has double peaks at around 4.5 and 4.7\,$\mu$m.
We attribute the emission to highly blue-shifted ($-3000 \sim -6000$\,km\,s$^{-1}$)
fundamental ro-vibrational transitions ($v=1-0$) of CO molecules of temperatures
of 1200--3700\,K with red continuum.  

We discuss two possibilities of the nature of $S2$: outflow associated with $S1$ and supernova
ejecta.  Detection of the CO fundamental ro-vibrational emission toward OMC-1 supports the
association of $S2$ with outflow from the massive YSO $S1$, although the characteristics of the CO emission
are different.
From simple comparison with the {\it ISO}/SWS spectra of OMC-1,  non-detection of
H$_2$ in $S2$ suggests a more H$_2$-poor environment in $S2$ than in OMC-1.
It also seems to be difficult to reconcile the large blue-shift velocity with the outflow origin.
The strong similarity of the $S2$ spectra to those toward the ejecta knots in the Cas A SNR suggests
that $S2$ may be associated with SN ejecta.  EGO G318.05+0.09 is located within
the SNR G318.2+0.1, however, there is no supporting evidence for infrared emission associated with this SNR.
There is no other evidence to relate $S2$ with SN ejecta.
Observations with high spectral resolution and better quality are definitely needed to draw a clear conclusion
on the nature of $S2$.

The present observation suggests that the greenish color of EGOs
has various origins and that one EGO could contain more than one objects of different types.  It demonstrates
significance of infrared spectroscopy for the study of the nature of individual objects.  There might
be similar CO emission associated in EGOs.

\acknowledgements
The authors thank the referee W. Reach for very useful comments, which greatly improved the paper.
This work is based on observations with {\it AKARI}, a JAXA project
with the participation of ESA.  The authors thank all the members of the {\it AKARI} project
for their continuous support.  They would also 
like to thank Fumihiko Usui for his help in the data reduction,
Bon-Chul Koo for the information on the SNR G318.2+0.1, and Fran\c{c}ois Boulanger for
useful discussion.  This research has made use of the NASA/IPAC Infrared Science Archive, 
which is operated by the Jet Propulsion Laboratory, California Institute of Technology, 
under contract with the National Aeronautics and Space Administration.
The laboratory data of the H$_2$O ice are obtained from the Leiden atomic and molecular database.
TM received financial support from a Grant-in-Aid for JSPS Fellows.
AA was supported by the University of Tokyo Research Internship
Program (UTRIP) of 2014, which initiated this work.
This work is supported by Grants-in-Aid
for Scientific Research from the Japan Society for the Promotion of Science
(nos. 23103004 and 16H00934).

\vspace{5mm}
\facilities{AKARI, Spitzer}


\listofchanges

\end{document}